\documentclass{emulateapj}
\usepackage{apjfonts}

\newcommand{\mum}{\ifmmode{\rm \mu m}\else{$\mu$m}\fi}

\begin{document}

\title{The distribution of silicate strength in Spitzer spectra of AGNs and ULIRGs}

\author{
Lei~Hao\altaffilmark{1},
D.~W.~Weedman\altaffilmark{1},
H.~W.~W.~Spoon\altaffilmark{1,2},
J.~A.~Marshall\altaffilmark{1},
N.~A.~Levenson\altaffilmark{3},
M.~Elitzur\altaffilmark{3},
J.~R.~Houck\altaffilmark{1}
%L.~Armus\altaffilmark{3},
\email{haol@isc.astro.cornell.edu}
}
\altaffiltext{1}{Cornell University, Astronomy Department, 
  Ithaca, NY 14853-6801}
\altaffiltext{2}{Spitzer Fellow}
\altaffiltext{3}{Departement of Physics and Astronomy, University of Kentucky, Lexington, KY 40506}
%\altaffiltext{3}{Caltech, Spitzer Science Center, MS 220-6, Pasadena, CA 91125}

\begin{abstract}

A sample of 196 AGNs and ULIRGs observed by the Infrared Spectrograph (IRS) on $Spitzer$ is analyzed to study the distribution of the strength of the 9.7\mum\ silicate feature.  Average spectra are derived for quasars, Seyfert 1 and Seyfert 2 AGNs, and ULIRGs. We find that quasars are characterized by silicate features in emission and Seyfert 1s equally by emission or weak absorption.  Seyfert 2s are dominated by weak silicate absorption, and ULIRGs are characterized by strong silicate absorption (mean apparent optical depth about 1.5).  Luminosity distributions show that luminosities at rest frame 5.5\mum\  are similar for the most luminous quasars and ULIRGs and are almost 10$^{5}$ times more luminous than the least luminous AGN in the sample. The distributions 
of spectral characteristics and luminosities are compared to those of optically faint infrared sources at z $\sim$ 2 being discovered by the IRS, which are also characterized by strong silicate absorption. It is found that local ULIRGs are a similar population, although they have lower luminosities and somewhat stronger absorption compared to the high redshift sources. 

\end{abstract}


%%%%%%%%%%%%%%%%%%%%%%%%%%%%%%%%%%%%%%%%%%%%%%%%%%%%%%%%%%%%%%%%%%%%%%%%%%

\section{Introduction}

Previous observations with the Infrared
Spectrograph (IRS)\footnote{The IRS was a collaborative venture between Cornell
University and Ball Aerospace Corporation funded by NASA through the
Jet Propulsion Laboratory and the Ames Research Center.} on $Spitzer$ \citep{hou_etal_04} have discovered a high redshift population of optically faint infrared sources.  This population has been derived from sources having extreme infrared to optical flux ratios (IR/opt $>$ 40, IR/opt = $\nu$f$_{\nu}$(24\mum)/$\nu$f$_{\nu}$($R$)) as selected from surveys at 24\mum\  with the Multiband Imaging Photometer \citep[MIPS,][]{rie_etal_04} on $Spitzer$. Redshifts were determined from mid-infrared spectral features for 43 of 58 sources observed in the Bo\"{o}tes field \citep{hou_etal_05,wee_etal_06_boote} and for 14 of 18 radio sources in the $Spitzer$ First Look Survey (FLS) field \citep{wee_etal_06_radio}. For the 57 sources which have redshifts, the redshift (z) ranges from 0.7 to 2.7, with a median z = 2.2.  53 of these 57 sources have silicate absorption, with only 4 having redshifts derived from polycyclic aromatic hydrocarbon (PAH) emission. This is in contrast to spectral features of sources selected on the basis of their near-infrared or submillimeter characteristics of the spectral energy distributions, most of which show PAH emission features \citep{yan_etal_05, lut_etal_05, wee_etal_06_swire}.
 
The presence of strong silicate absorption and the absence of PAH features in samples
chosen from extreme values of IR/opt have been interpreted to mean that
these sources are heavily obscured AGN. Extreme IR sources with confirmed optical AGN properties were also found by ISOCAM \citep{haa_etal_04_agn}. It is therefore essential to
compare the characteristics of these high redshift, heavily absorbed
sources to known examples of AGNs and Ultraluminous Infrared Galaxies
(ULIRGs) to determine if there are local analogues to the obscured, high
redshift sources. In this letter, we examine this issue by studying the distribution of silicate strengths of local AGNs and ULIRGs. 

The distribution of the mid-IR properties of different types of known AGNs has been discussed in various studies \citep{hec_etal_94, haa_etal_03, sie_etal_04, cla_etal_00, shi_etal_06, buc_etal_06}. The mid-IR spectral features, especially the silicate features at 9.7 and 18\mum\,  are particularly important in the context of the unification model \citep[eg.][]{antonu_93}. The unambiguous detection of silicate emissions in quasars with $Spitzer$ is a strong support for the unification model \citep{sie_etal_05, hao_etal_05_se}, but there is a large variety of silicate features in AGNs. Silicate absorption is seen in some type I AGNs \citep{wee_etal_05} and emission in some type II AGNs \citep{stu_etal_05}. It is important, therefore, to investigate the distribution of silicate strengths within a large sample of different types of AGNs. The distribution will constrain the models of AGNs \citep[eg.][]{efstat_06, fri_etal_06, dul_vbem_05, nen_etal_02, pie_kro_93, pie_kro_92} and in turn provide insights into the geometry of the dusty structure, the optical depth of the clouds, and the filling factor along the line of sight. 
%In section 2, we describe the selection of a large sample of local AGN and ULIRGs and their data analysis. In section 3, we present results for the distribution of their silicate strength. In section 4, we conclude with a discussion of its implications.

% understand the AGN obscuration at 10um vs the obscuration reflected by the broad-band photometry -- AGN extinction law.

%%%%%%%%%%%%%%%%%%%%%%%%%%%%%%%%%%%%%%%%%%%%%%%%%%%%%%%%%%%%%%%%%%%%%%%%%%

\section{Observations and data reduction}
  We gathered a large sample of local AGNs and ULIRGs from archival and published literature that have been observed with the {\it Spitzer} IRS. The AGN sample combines the unpublished GTO archival data (program 14) and published data from \cite{hao_etal_05_se}, \cite{wee_etal_05}, \cite{buc_etal_06} and \cite{shi_etal_06}. The ULIRG sample is drawn from the GTO program (program 105),\citep{arm_etal_06, spo_etal_06}. The total sample includes objects having a large variety of classifications: radio-quiet QSOs, radio-loud QSOs, Seyfert 1s and Seyfert 2s, IRAS-discovered sources, and 2MASS-selected red AGNs \citep{smi_etal_02}. 

We use only the IRS observations of sources which include all low-resolution modules: Short-Low2 (SL2, 5.2-7.7\mum), Short-Low1 (SL1,7.4-14.5\mum), Long-Low2 (LL2, 14-21.3\mum) and Long-Low1 (LL1, 19.5-38\mum). We also exclude objects that have redshifts larger than 0.5, so that we can compare characteristics in similar rest frames.  

To define subsamples of objects, we adopt the optical classifications from \cite{ver_ver_06}, which include three catalogues: `quasars', `AGNs' and `Blazars'. In these catalogues, quasars are defined as objects that are starlike or with a starlike nucleus, broad emission lines, and brighter than absolute magnitude M$_{B}$=-23. AGNs include Seyfert 1s, Seyfert 2s and LINERs that are fainter than M$_{B}$=-23. Seyfert 1s are further divided into five subgroups: Seyfert 1.0, 1.2, 1.5, 1.8 and 1.9 based on the appearance of the Balmer lines. For our study, we compiled a quasar sample including all objects from the `quasar' catalogue that have been classified as S1, S1.2, S1.5 or S1n (Narrow Line Seyfert 1s) from their optical emission line properties; a Seyfert 1 sample from the `AGN' catalogue as sources that have optical spectroscopic classifications of S1, S1.2, S1.5 and S1n; and a Seyfert 2 sample from the `AGN' catalogue that have classifications of S2, S1h (Seyfert 2s with broad Balmer lines in the polarized light) and S1i (Seyfert 2s with broad Pa $\beta$ in the near IR). Objects that are classified as S1.8 or S1.9 have weak broad Balmer lines, and their identifications as Seyfert 1s depends very much on the signal-to-noise of the spectra. Therefore, we do not include them in our Seyfert 1 sample. We also do not consider objects classified as LINERs in our sample. 

Our ULIRG sample includes all objects observed in program 105 with all four low resolution modules and having $z<0.5$. It is a loosely defined sample, chosen primarily on the basis of bolometric infrared luminosity \citep{arm_etal_06}. Some ULIRGs also show optical characteristics of quasars, Seyfert 1s or Seyfert 2s according to \cite{ver_ver_06}, and these are also included in the three AGN samples. The final sample includes 24 quasars (including 3 ULIRGs), 45 Seyfert 1s (including 7 ULIRGs), 47 Seyfert 2s (including 8 ULIRGs), and 98 ULIRGs, in total 196 separate sources. 

The spectra were extracted using the SMART analysis package \citep{hig_etal_04}.  Extractions were done differently for observations done in {\it staring} mode and in {\it mapping} mode. For {\it staring} mode observations, we extract spectra from the DROOP products provided by the Spitzer Science Center in Pipeline version 11.0.2 and 14.0, and subtract background by differencing the SL1 and SL2 or LL1 and LL2. The spectra are then calibrated using the IRS standard stars HD173511 (5.2-19.5\mum) and $\xi$Dra(19.5-38.5\mum). The {\it mapping} mode spectra are extracted using the default column extraction in SMART, utilizing the pipeline BCD products and extracting only the central image of each observation after subtracting the outermost image as background (see \cite{buc_etal_06} for technical details). The final spectra were stitched and scaled to the LL1 spectrum.
 
% The method to fit the continuum and measure the silicate depth -- See Spoon et al., 2006. (slightly explain here, later)   

%The paper includes about 250 AGN and ULIRG spectra. The photometry study include a subset of the sample that have iras, 2mass and sdss photometry (or B band photometry) 

%%%%%%%%%%%%%%%%%%%%%%%%%%%%%%%%%%%%%%%%%%%%%%%%%%%%%%%%%%%%%%%%%%%%%%%%%%
\section{Results}

\begin{figure}[t!]
\centerline{
\includegraphics[angle=0.,width=\hsize]{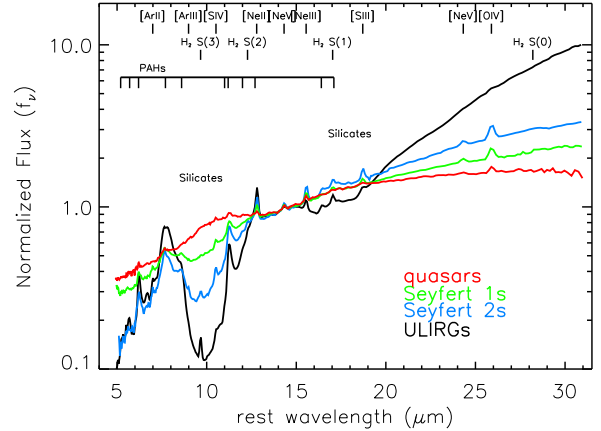}}
\caption{The average spectra of QSOs (red), Seyfert 1s (green), Seyfert 2s (blue) and ULIRGs (black). The average spectra are obtained by averaging after first nomalizing each spectra at 14.5\mum\  in the rest frame. The 10\mum\ silicate strength measured from the average spectra are: +0.20, -0.21, -0.54, and -1.44 for quasars, Seyfert 1s, Seyfert 2s and ULIRGs, respectively.
\label{average}}
\end{figure}

The mid-IR spectra of the objects in the four samples show great variety. The average spectra of the four samples are presented in Figure~\ref{average}\footnote{The
average spectra are available from \url{http://}, which will be completed after the paper's been accepted}. The average quasar spectrum clearly shows silicate features in emission at both 10 and 18\mum. There are very weak indications of PAH emission at 6.2, 7.7 and 11.3\mum, and weak emission lines from [SIV]10.5\mum, [NeII]12.8\mum, [NeV]14.3\mum, [NeIII]15.5\mum, [SIII]18.7\mum, [NeV]24.3\mum\  and [OIV]25.9\mum. PAH emission is also detected in the average spectrum of PG quasars, even for those that do not show PAHs in their individual spectra \citep{schwei_etal_06}. For the Seyfert 1 average spectrum, it is not clear if the 10\mum\ silicate feature is in emission or absorption. All the PAH features and atomic emission lines are stronger compared with the quasars. The average spectrum of Seyfert 2s indicates even stronger PAH emission features. The silicate feature at 10\mum\  is clearly in absorption. The H$_{2}$ lines are stronger, but the atomic lines are similar to those in the Seyfert 1s. Stronger PAHs in Seyfert 2s compared with Seyfert 1s are also seen in \cite{buc_etal_06}. The spectrum of the average ULIRG shows the strongest PAH emission compared with the AGNs and also shows clear absorption at both the 10 and 18 \mum\  silicate features. The H$_{2}$ lines are also more pronounced than those in AGNs. The mid-IR spectral slope decreases significantly from ULIRGs to quasars.  

%%%%%%%%%%%%%%%%%%%%%%%%%%%%%%%%%%%%%%%%%%%%%%%%%%%%%%%%%%%%%%%%%%%%%%%%%
%%%%%%%%%%%%%%%%%%%%%%%%%%%%%%%%%%%%%%%%%%%%%%%%%%%%%%%%%%%%%%%%%%%%%%%%%

To quantify the silicate strength at $\sim$ 10\mum, we define the silicate strength $S_{10}$ as:
\begin{equation}
  S_{10}=\ln {{f_{obs}(10\mum)}\over{f_{cont}(10\mum)}},
\end{equation}
where $f_{obs}(10\mum)$ is the observed flux density at the peak of the 10\mum\  feature, and $f_{cont}(10\mum)$ is the continuum flux at the peak wavelength, extrapolated above the silicate absorption feature (or below the emission feature). For determination of $f_{cont}(10\mum)$, we adopt the methods described in \cite{spo_etal_06}. This uses three different techniques for PAH-weak spectra, PAH-dominated spectra, and absorption-dominated spectra to avoid the contamination of various PAH emission features to the continuum and to make best use of uncontaminated continuum regions.  
%The flux densities are measured at a fixed wavelength of 9.7\mum, although it should be noted that the majority of prominent silicate emission features have peak wavelengths longer than 9.7\mum, as can be seen in the average spectra. Such a shift of the peak wavelength may indicate an increased average grain size or presence of crystalline silicate. Most of the silicate absorption features do show the minimum close to 9.7\mum.

\begin{figure}[th!]
\centerline{
\includegraphics[angle=0.,width=\hsize]{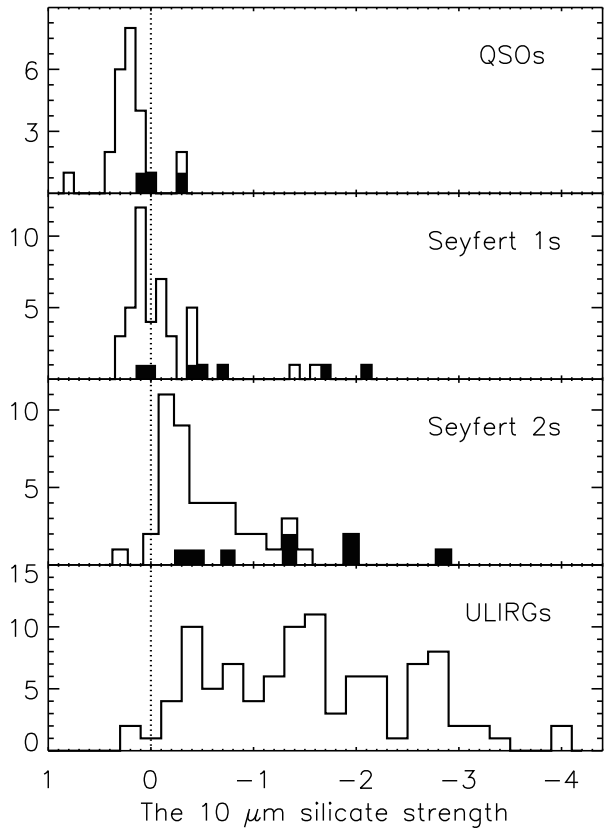}}
\caption{The distribution of the 10\mum\ silicate strength, $S_{10}$, for QSOs, Seyfert 1s, Seyfert 2s and ULIRGs. Silicate absorption increases to the right. In QSOs, Seyfert 1s and Seyfert 2s, the distributions of sources that are also in the ULIRG sample are shaded. The averages of the 10\mum\ silicate strengths are 0.20 for quasars, -0.18 for Seyfert 1s, -0.61 for Seyfert 2s and -1.56 for ULIRGs. The range of the silicate strengths are $0\le S_{10}\le 0.4$, $-0.7\le S_{10}\le 0.3$, $-1.6\le S_{10}\le 0$, $-4\le S_{10}\le 0.2$ for most quasars, Seyfert1, Seyfert 2s and ULIRGs.
\label{sildist}}
\end{figure}

% Within each sample, the mid-IR spectra span a large variety. Here we focus on the silicate features and in Figure 2, we show the distribution of the 10 \mum\  silicate feature strengths in the four samples. 

In Figure 2, we show the distribution of the 10\mum\  silicate strengths of quasars, Seyfert 1s, Seyfert 2s and ULIRGs. For objects with silicate strength close to 0 (the bin crossing $S_{10}=0$ in Figure 2), identification of the feature as emission or absorption is strongly affected by the S/N of the spectra and the uncertainty of the continuum determination. We ignore the objects in this bin for further statistical analysis. The majority of quasars have the 10\mum\ silicate feature in emission, but 2 cases clearly show the feature in absorption. These include one ULIRG (IRAS 00275-2859) and a source from the 2MASS-selected red AGN sample (2MASSi J1258074+232921) \citep{shi_etal_06}. Their optical spectra clearly show broad Balmer emission and blue continuum \citep{zhe_etal_02, smi_etal_02}, which do not distinguish them from typical quasars. But the fact that they are selected in the IRAS or 2MASS red AGN sample which select for dusty sources agrees with their having silicate features in absorption.  For quasars that have silicate emission, the features are weak. The strongest silicate strength in quasars has a value of $\sim$ +0.75 (PG1351+640), but the remainder all have silicate strength less than 0.4. The average silicate strength in the quasar sample is 0.20. 

In the Seyfert 1 sample, even more objects show silicate absorption at 10 \mum. Excluding objects that have measured silicate strength close to 0, the number of objects that have silicate in emission is about the same as the number of objects that have silicate absorption. All of those in emission have strengths less than 0.35, and the average silicate strength in the sample is $-0.18$. There are four Seyfert 1 sources that show silicate absorption deeper than $-1$. They are 2MASSiJ1307006+233805, 2MASSiJ1453315+135358, IRAS14348-1447 and UGC5101. However, their optical spectra do not show unambiguous broad Balmer lines \citep{smi_etal_02, vei_etal_95, san_etal_88}, and it is possibly arbitrary to classify them as Seyfert 1s. The deepest silicate absorption for sources with unambiguously identified broad Balmer lines is $-0.65$, from Mrk231 \citep{wee_etal_05}. We particularly note that the silicate strength in quasars is more dominated by emission than in Seyfert 1 AGNs, and that the luminosities of the quasars as shown in Figure 4 are systematically greater than for Seyfert 1s.

Almost all of the Seyfert 2s in the sample show the 10 \mum\  silicate feature in absorption except for one case: IRAS~F01475-0740. The average silicate strength of the sample is -0.61. Silicate emission is also detected in the average spectra of type II QSOs \citep{stu_etal_06}, although IRAS F01475-0740 in our sample is not a highly luminous source.

The ULIRG sample is dominated by sources with strong silicate absorption with 10\mum\ silicate strength $-4\le S_{10}\le 0.2$. There are 2 cases of weak silicate emission: Mrk1014 and IRAS 07598+6508, but their optical spectra place the former in the quasar and the latter in the Seyfert 1 samples. The average of the 10\mum\  strength in the ULIRG sample is $-1.56$.

In Figure 3, we show the comparison between the infrared spectral slope and the silicate strength. There is a clear correlation of the silicate strengths with the IR slopes as measured from the flux ratio of 14.5\mum\  to 27.5\mum, a result that is also evident in Figure 1.  But there is less correlation of the silicate strength with the slope measured from the flux ratio at 5.5\mum\ and 14.5\mum\ or 5.5\mum\ and 27.5\mum\ . This suggests that the silicate feature arises primarily within the cooler dust responsible for the mid-IR emission of deeply absorbed ULIRGs, and that the near-IR emission does not arise from the same dust. \citet{lev_etal_06} suggest that absorption features as strong as those observed require a nuclear source to be deeply embedded in a smooth distribution of dusty material that is both geometrically and optically thick.  Given their extreme obscuration, the near-IR emission would be heavily attenuated. The observed near-IR emission would thus have to have a different origin, and not fill in the deep silicate absorption trough. 

\begin{figure}[t!]
\centerline{
\includegraphics[angle=90.,width=\hsize]{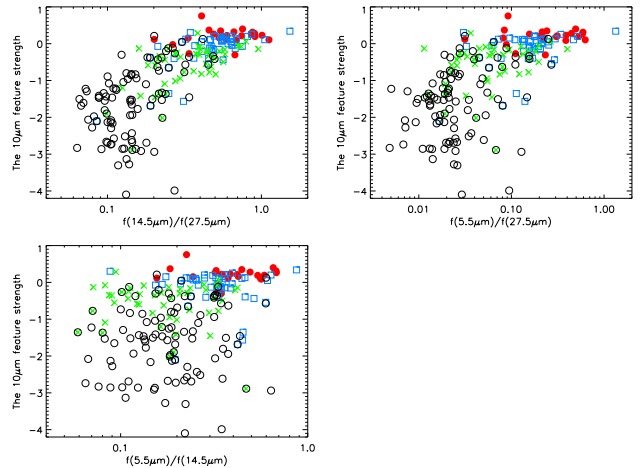}}
\caption{The correlation between the observed strength of the silicate feature, $S_{10}$, and various IR slope. The open circles are ULIRGs, squares are Seyfert 1s, crosses are Seyfert 2s and filled circles are quasars. The color code is the same as in Figure 1.
\label{tau_irslope}}
\end{figure}

\begin{figure}[th!]
\centerline{
\includegraphics[angle=0.,width=\hsize]{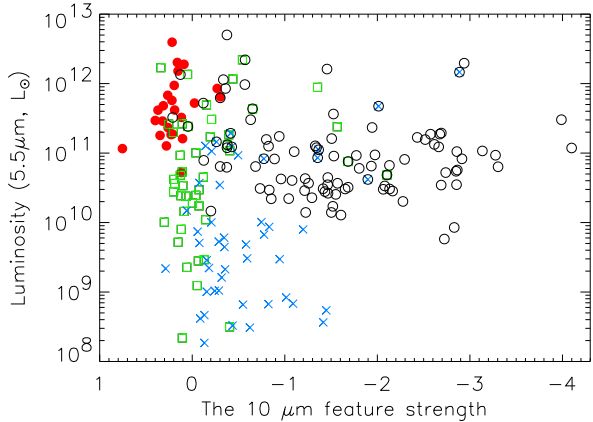}}
\caption{The correlation between the observed strength of the silicate feature, $S_{10}$, and the IR luminosity at 5.5\mum\ . The symbols and color code are as in Figure 3; silicate absorption increases to the right.
\label{tau_irlum}}
\end{figure}

%%%%%%%%%%%%%%%%%%%%%%%%%%%%%%%%%%%%%%%%%%%%%%%%%%%%%%%%%%%%%%%%%%%%%

\section{Discussion}

Our sample is not complete in any sense, because it includes a variety of objects selected with different criteria.  However, with 196 objects, this is the largest available dataset of local AGNs and ULIRGs observed by the  IRS which enables us to measure the distribution of the silicate strength. The distribution of the silicate strengths should be representative of the different categories of objects in the local universe.  

The results in Figure 2 are important for understanding the faint sources (f$_{\nu}$(24\mum) $\sim$ 1 mJy) at high redshift which have large values of IR/opt for which S/N is poor and features must be strong for a redshift to be assigned.  Of the 53 sources with redshifts derived from silicate absorption as reported in \citet{hou_etal_05}, \citet{wee_etal_06_radio}, and \citet{wee_etal_06_boote}, 44 sources had silicate absorption at least as deep as in Mrk 231, for which the silicate strength is -0.7 (the absorbed continuum is at a level of 50\% of the continuum extrapolated above the absorption).  16 of 76 sources observed (21\%) had no apparent absorption, to a limit for $S_{10}$ of $\sim$ -0.3. Figure 2 indicates that only 10\% of ULIRGS have silicate absorption shallower than $-0.3$ ($S_{10} \ge -0.3$), implying that ULIRGS are even more absorbed than the faint sources characterised by extreme IR/opt, although the absence of features in the optically faint sources could sometimes arise because features are redshifted out of the IRS spectral range. 
%However, the Seyfert 1, Seyfert 2, and quasar distributions show systematically weaker absorption than the high redshift sources, so some mixture of sources with ULIRG characteristics and AGN characteristics could explain the absorption distribution in the high-redshift sources. 

Figure 4 shows the 5.5\mum\  luminosity compared with the 10\mum\  silicate feature strength for quasars, Seyfert 1s, Seyfert 2s and ULIRGs. It is particularly useful for interpreting the sources at high redshift, which must be extremely luminous in order to fall within samples having f$_{\nu}$(24\mum) $\sim$ 1 mJy and z $\sim$ 2.  For example, the most luminous source yet found among these samples is source \#9 in \citet{hou_etal_05}, which has luminosity of $7\times 10^{12}$L$_\odot$ at 5.5\mum, and exceeds any source in our current sample.  All of the sources at z $\sim$ 2 reported so far have luminosities above $\sim 2\times 10^{12} L_\odot$ at rest wavelength of $\sim 6\mum$.  Figure 4 indicates that both quasars and ULIRGS can have luminosities that match these high redshift sources.  High redshift quasars have been discovered in the $Spitzer$ 24\mum\  samples, but these quasars have $R<21.7$ \citep{bro_etal_06} and therefore have much brighter optical magnitudes than the obscured sources with extreme IR/opt. This means that examples at high redshift similar to the "quasars" in Figure 4 have been discovered in the infrared, but they are not heavily attenuated in the optical. This implies that the local analogues to the optically faint sources at high redshift would be those luminous ULIRGs in Figure 4. 

The nature of the dust distribution in various types of AGN and the relation of this to a unification scheme is uncertain.  For example, some silicate emission from AGNs may arise on scales larger than the torus of the unification scheme \citep{stu_etal_05}. The distributions of the silicate strengths presented here are intended to give statistical meaning to understanding the dusty structure of AGNs and ULIRGs, rather than interpreting individual objects.  For example, the average spectra provide a practical means to identify fundamental differences among the nuclear dusty environments of the different families. A significant number of ULIRGs show very deep silicate absorption ($S_{10}<-2.5$), in strong contrast to the quasars of similar luminosity which show no obvious sign of absorption and also in contrast to the classical optically-classified AGNs. This reflects a different geometry of the dusty material in these deeply absorbed ULIRGs \citep{lev_etal_06}. The distribution also reveals a significant number of silicate absorptions in Seyfert 1s and a few in quasars, and one silicate emission in a Seyfert 2. These results are not expected from the simplest unification scenario, although several AGN models can accomodate such exceptions \citep[eg.][\ M. Nenkova et al., in preparation]{nen_etal_02, efstat_06}. The fraction of these exceptional cases and the ranges of their silicate strengths provide important constraints on these models. It should be noted that the fraction of Seyfert 1s having weak silicate absorptions is much larger than the fraction of Seyfert 2s with silicate emission, which requires a dust geometry that allows the broad-line region to be easily visible even when there is significant optical depth through the dust clouds themselves.

%\acknowledgments
We thank D. Devost, G. Sloan, and P. Hall for help in improving our IRS spectral analysis and M. Strauss for helpful discussions. This work is based on observations made with the Spitzer Space Telescope, which is operated by the Jet Propulsion Laboratory, California Institute of Technology, under NASA contract 1407. Support for this work by the IRS GTO team at Cornell University was provided by NASA through Contract Number 1257184 issued by JPL/Caltech. 

%%%%%%%%%%%%%%%%%%%%%%%%%%%%%%%%%%%%%%%%%%%%%%%%%%%%%%%%%%%%%%%%%%%%%

%\acknowledgments Support for this work was provided by NASA through
%Contract Number 1257184 issued by the Jet Propulsion Laboratory,
%California Institute of Technology under NASA contract 1407.  HWWS was
%supported under this contract through the Spitzer Space Telescope
%Fellowship Program.

%%%%%%%%%%%%%%%%%%%%%%%%%%%%%%%%%%%%%%%%%%%%%%%%%%%%%%%%%%%%%%%%%%%%%

%\bibliographystyle{/home/haol/papers/reference/aj.bst}
%\bibliography{/home/haol/papers/reference/allref_hao}

\end{document}